\documentstyle[epsfig,amstex,amssymb,iopconf1]{article}
\newcommand{\zpc}[3]{Z.\ Phys.\ {\bf C#1} (19#2) #3}
\newcommand{\rmp}[3]{Rev.\ Mod.\ Phys.\ {\bf C#1} (19#2) #3}
\newcommand{\plb}[3]{Phys.\ Lett.\ {\bf B#1} (19#2) #3}

\newcommand{\npb}[3]{Nucl.\ Phys.\ {\bf B#1} (19#2) #3}
\newcommand{\prd}[3]{Phys.\ Rev.\ {\bf D#1} (19#2) #3}

\newcommand{\prl}[3]{Phys.\ Rev.\ Lett.\ {\bf #1} (19#2) #3}

\def\simgt{\rlap{\lower 3.5 pt \hbox{$\mathchar \sim$}} \raise 1pt \hbox {$>$}}
\def\simlt{\rlap{\lower 3.5 pt \hbox{$\mathchar \sim$}} \raise 1pt \hbox {$<$}}

\newcommand{\beq}{\begin{equation}}
\newcommand{\eeq}{\end{equation}}
\newcommand{\bea}{\begin{eqnarray}}
\newcommand{\eea}{\end{eqnarray}}

\newcommand{\alps}{\mbox{$\alpha_{\mbox{\scriptsize s}}$}}
\newcommand{\alpsq}{\mbox{$\alpha_{\mbox{\scriptsize s}}^{2}$}}

\newcommand{\LQ}{L\!Q}
\newcommand{\LQb}{\overline{L\!Q}}
\newcommand{\Slqb}{\overline{S}}
\newcommand{\Vlqb}{\overline{V}}
\newcommand{\mlq}{\mbox{$m_{\scriptstyle \LQ}$}}
\newcommand{\msq}{\mbox{$m^2_{\scriptstyle \LQ}$}}
\newcommand{\ms}{\mbox{$m_{\scriptstyle S}$}}

\newcommand{\mg}{\mbox{$m_{\tilde{g}}$}}

\catcode`\@=11
\def\@citex[#1]#2{\if@filesw\immediate\write\@auxout{\string\citation{#2}}\fi
  \def\@citea{}\@cite{\@for\@citeb:=#2\do
    {\@citea\def\@citea{,\penalty\@m}\@ifundefined
       {b@\@citeb}{{\bf ?}\@warning
       {Citation `\@citeb' on page \thepage \space undefined}}%
\hbox{\csname b@\@citeb\endcsname}}}{#1}}
\def\citer{\@ifnextchar [{\@tempswatrue\@citexr}{\@tempswafalse\@citexr[]}}
\def\@citexr[#1]#2{\if@filesw\immediate\write\@auxout{\string\citation{#2}}\fi
  \def\@citea{}\@cite{\@for\@citeb:=#2\do
    {\@citea\def\@citea{--\penalty\@m}\@ifundefined
       {b@\@citeb}{{\bf ?}\@warning
       {Citation `\@citeb' on page \thepage \space undefined}}%
\hbox{\csname b@\@citeb\endcsname}}}{#1}}
\begin{document}

\begin{flushright}
RAL-TR-97-033\\
July 1997
\end{flushright}

\title{Leptoquark production at the Tevatron\footnote{Talk presented at
       ``Beyond the Desert 97'' -- Accelerator and Non-Accelerator
       Approaches, Castle Ringberg, Tegernsee, Germany, 8-14 June 1997,
       to appear in the proceedings}}

\vspace*{-5mm}

\author{Michael Kr\"amer\footnote{E-mail: Michael.Kraemer@rl.ac.uk}}

\affil{Rutherford Appleton Laboratory, Chilton, OX11 0QX, UK}

\beginabstract
I discuss the production of leptoquarks in $p\bar{p}$ collisions at
the Fermilab Tevatron.  Evaluated on the basis of a recent
next-to-leading order calculation, the combined \mbox{D0} and
\mbox{CDF} leptoquark searches lead to a parameter-free lower mass
bound of about $\ms \;\simgt\; 240$~GeV for scalar leptoquarks or
squarks decaying solely to a first generation charged lepton plus
quark. These results provide very stringent constraints on any
explanation of the recently observed \mbox{HERA} excess events in
terms of new particle production.
\endabstract

%%%%%%%%%%%%%%%%%%%%%%%%%%%%%%%%%%%%%%%%%%%%%%%%%%%%%%%%%%%%%%%%%%%%%%%%

\section{Introduction\label{sec-intro}} 
% ----------------------------------- %

The recent observation of an excess of events in deep-inelastic
positron-proton scattering at \mbox{HERA} \cite{HERA-97} has aroused
new interest in physics beyond the Standard Model.  Theoretical
speculations have focussed on two different scenarios to explain the
experimental results: contact interactions at an effective scale
$\Lambda_{eq} \;\simgt\; 1.5$~TeV, and narrow resonance formation at a
mass scale $M\sim 200$~GeV. The resonance interpretation is based in
particular on the \mbox{H1} data in \cite{HERA-97} which appear to
cluster in a narrow range at invariant $(eq)$ masses of about 200~GeV.
Such resonances can be identified with scalar squarks in
supersymmetric theories with $R$-parity breaking or leptoquarks in
general \cite{DRR-97}. If low-mass leptoquark-type states exist, they
are produced at significant rates via QCD interactions in $p\bar{p}$
collisions at the Fermilab Tevatron.  Evaluated on the basis of a
recent next-to-leading order calculation \cite{KPSZ-97}, the \mbox{D0}
\cite{D0-97} and \mbox{CDF} \cite{CDF-97} leptoquark searches lead to
a parameter-free lower mass bound of about $\ms \;\simgt\; 240$~GeV
for scalar leptoquark-type states decaying solely to a first
generation charged lepton plus quark. The corresponding mass bound on
vector leptoquarks appears significantly higher, even if the unknown
anomalous couplings of vector leptoquarks to gluons are chosen such as
to minimize the cross section \cite{BBK-97}.  These results rule out
the standard leptoquark interpretation of the \mbox{HERA} data as
discussed in Section~\ref{sec-hera} and provide very stringent
constraints on any explanation of the excess events in terms of new
particle production.

The article is organized as follows: In Section~\ref{sec-hera}, I
briefly review the leptoquark interpretation of the \mbox{HERA} data.
Single and pair production of scalar and vector leptoquarks in
$p\bar{p}$ collisions at the Fermilab Tevatron are discussed in
Section~\ref{sec-tev}.  The next-to-leading order (NLO) QCD
corrections to scalar-leptoquark pair production in $p\bar{p}$
collisions have been evaluated recently. As shown in
Section~\ref{sec-nlo}, including the higher-order contributions
stabilizes the theoretical prediction and increases the size of the
cross section for renormalization/factorization scales close to the
mass of the leptoquarks. The leptoquark searches at the Fermilab
Tevatron and the resulting mass bounds for leptoquarks and scalar
squarks in supersymmetric theories with $R$-parity breaking are
described in Section~\ref{sec-bounds}. I conclude in
Section~\ref{sec-con}.

\section{Leptoquarks: basic set-up\label{sec-hera}}
% ---------------------------------------------------------- %
The interactions of leptoquarks with lepton-quark pairs can be
described by an effective low-energy lagrangian including the most
general dimensionless and \mbox{SU(3) $\times$ SU(2) $\times$ U(1)}
invariant couplings \cite{BRW-87}.  The allowed states are classified
according to their spin ($S=0,1$), weak isospin and fermion number
($F=0,-2$). The leptoquark-fermion couplings of low-mass leptoquark
states are severely constrained by low-energy experiments. The
couplings are assumed to be baryon- and lepton number conserving in
order to avoid rapid proton decay and family diagonal to exclude FCNC
beyond CKM mixing.  Moreover, the leptoquark-fermion couplings have to
be essentially chiral to preserve the helicity suppression in leptonic
pion decays.

Additional constraints on the leptoquark-fermion couplings from the
\mbox{HERA} $(e^-p)$ data and atomic parity violation measurements
only allow certain fermion number zero scalar and vector leptoquarks
as a possible source of the $(e^+p)$ excess. Within the standard
scheme discussed above, {\em i.e.}\ imposing chiral and family
diagonal couplings to evade the low-energy bounds, and assuming the
fermionic content of the Standard Model, all leptoquark candidates
consistent with the \mbox{HERA} data decay with 100\% branching
fraction to a first generation charged lepton plus quark.

\section{Leptoquark production at the Fermilab Tevatron\label{sec-tev}}
% ------------------------------------------------------------------- %
The most stringent mass limits on leptoquark-type states arise from
direct searches in $p\bar{p}$ collisions at the Fermilab Tevatron.
The leptoquark cross section is dominated by pair production
\begin{equation}\label{eq-lqlqb}
p + \bar{p} \to \LQ + \LQb + X 
\end{equation} 
which proceeds through quark-antiquark annihilation and gluon-gluon
fusion. The interactions of scalar leptoquarks with gluons are
completely determined by the non-abelian SU(3)$_C$ gauge symmetry of
scalar QCD so that the theoretical predictions for the pair production
of scalar leptoquarks are parameter-free. Vector leptoquarks can have
additional anomalous couplings $(\kappa_V,\lambda_V)$ to gluons which
violate unitarity in the production cross section. These couplings
vanish in any theory wherein vector leptoquarks appear as fundamental
gauge bosons of an extended gauge group, in the absence of a definite
model however the general case $\kappa_V,\lambda_V \neq 0$ should be
considered.

\noindent
The partonic cross sections that contribute to leptoquark pair
production (\ref{eq-lqlqb}) are given by
\begin{alignat}{3}\label{eq-ssb}
&\hat{\sigma}_{\mbox{\scriptsize LO}}[\,q\bar{q}\to S\Slqb\,] 
&\;=\;&
\frac{\alpsq\pi}{27\hat{s}}\,2\beta^3 \hfill \\
&\hat{\sigma}_{\mbox{\scriptsize LO}}[\,gg\to S\Slqb\,] 
&\;=\;&
\frac{\alpsq\pi}{96\hat{s}}\,
\left[ \beta \left( 41 - 31 \beta^2 \right)
%\right.\nonumber\\[1mm] 
%&&&\left.
       + \left( 18 \beta^2 - \beta^4 - 17 \right) 
         \ln\frac{1+\beta}{1-\beta}
\right]\nonumber\\
\intertext{for scalar leptoquarks \cite{GM-82} and} 
%\end{alignat}
%\begin{alignat}{3}
&\hat{\sigma}_{\mbox{\scriptsize LO}}[\,q\bar{q}\to V\Vlqb\,] 
&\;=\;&
\frac{\alpsq\pi}{27\hat{s}}\,\frac{\beta^3}{1-\beta^2}\,
\left[ 23 - 3\beta^2+\frac{4}{1-\beta^2}\right] \hfill \nonumber\\
&\hat{\sigma}_{\mbox{\scriptsize LO}}[\,gg\to V\Vlqb\,] 
&\;=\;&
\frac{\alpsq\pi}{24\hat{s}}\frac{1}{(1-\beta^2)}\,
\left[ \beta \left(\frac{523}{4} - 90 \beta^2 
+\frac{93}{4}\beta^4\right) \right. \nonumber\\
& & & \quad\quad\quad\left.
       - \frac34\left( 65 -83\beta^2 + 19\beta^4 - \beta^6 \right) 
         \ln\frac{1+\beta}{1-\beta}
\right]\label{eq-vvb}
\end{alignat}
for vector leptoquarks with vanishing anomalous couplings
$\kappa_V,\lambda_V = 0$ \cite{AW-86}. The invariant energy of the
subprocess is denoted by $\sqrt{\hat{s}}$ and $\beta =
(1-4\msq/\hat{s})^{\frac12}$ is the velocity of the produced
leptoquarks in their centre-of-mass system.  Results for the general
case of vector leptoquark pair production with $\kappa_V,\lambda_V
\neq 0$ are given in \cite{BBK-97}.  Contributions to the
quark-antiquark annihilation cross sections from lepton exchange in
the $t$-channel involve the leptoquark-fermion coupling and can be
neglected compared to the ${\cal{O}}(\alpsq)$ processes listed above.

\begin{figure}[ht]
\begin{center}
\epsfig{file=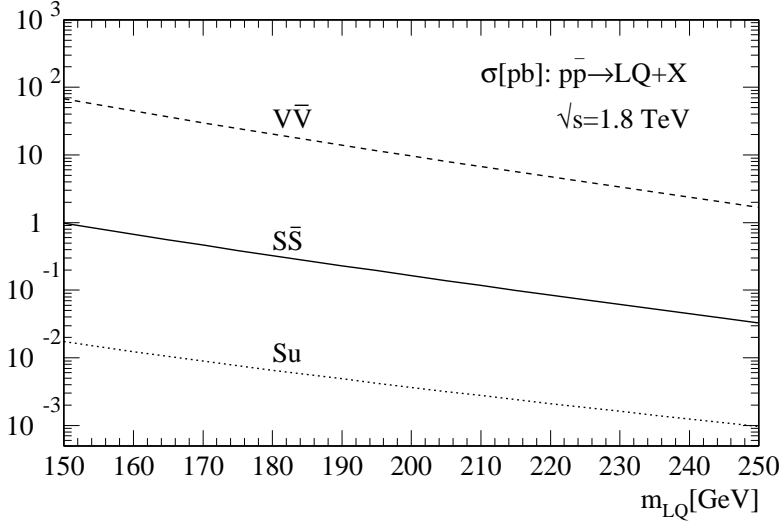,width=10.5cm}
\end{center}
\vspace*{-5mm}
\caption[dummy]{\it The leading-order cross sections for single 
  and pair production of scalar and vector leptoquarks
  $(\kappa_V,\lambda_V = 0)$ in $p\bar{p}$ collisions at the Tevatron
  as a function of the leptoquark mass $\mlq$. The CTEQ4L parton
  densities \cite{CTEQ-97} have been adopted and the
  renormalization/factorization scale has been set to $\mu=\mlq$.
  `$Su$' denotes the contribution to single scalar leptoquark
  production from scattering off $u$ and $\bar{u}$ quarks in the
  proton, taking $\lambda = e/10$.}
\label{fig-lo}
\end{figure}

The $p\bar{p}$ cross section is found by folding the parton cross
sections (\ref{eq-ssb},\ref{eq-vvb}) with the gluon and light--quark
luminosities in $p\bar{p}$ collisions. Due to the dominating
$q\bar{q}$ luminosity for large parton momenta, the total cross
section is built up primarily by quark-antiquark initial states for
leptoquark masses $\mlq\;\simgt\; 100$~GeV. From Figure~\ref{fig-lo}
one can infer that the leading order (LO) cross sections for $\mlq
\sim 200$~GeV scalar and vector leptoquarks are $\sigma[\,p\bar{p}\to
S\Slqb\,] \sim 0.2\,\mbox{pb}$ and $\sigma[\,p\bar{p}\to V\Vlqb\,]
\sim 10\,\mbox{pb}$ respectively, assuming vanishing anomalous
couplings of vector leptoquarks to gluons.  Even if the anomalous
couplings are chosen such as to minimize the cross section, the total
rate for vector leptoquark pair production is still a factor of two
larger than that for scalar leptoquarks \cite{BBK-97}.

%%%%%%%%%%%%%%%%%%%%%%%%%%%%%%%%%%%%%%%%%%%%%%%%%%%%%%%%%%%%%%%%%%%%%%%
%%%%%%%%%%%%%%%%%%%%%%%%%%%%%%%%%%%%%%%%%%%%%%%%%%%%%%%%%%%%%%%%%%%%%%%%

Another source of leptoquarks at the Tevatron is associated production
$p\bar{p} \to \LQ + \bar{l} + X$ \cite{HP-88}.  The cross section for
this process depends quadratically on the leptoquark-fermion coupling
and is thus significantly smaller than for leptoquark pair production.
Taking $\lambda = e/10\approx 1/30$ for illustration, one finds
$\sigma[\,p\bar{p}\to S\bar{l}\,] \sim 4 \times 10^{-3}\,\mbox{pb}$
for a $\ms = 200$~GeV scalar leptoquark with couplings to $u$ and
$\bar{u}$ quarks (see Figure~\ref{fig-lo}), {\em i.e.}\ too small to
be experimentally accessible at the moment.

The LO cross sections for scalar leptoquark pair production coincide
with the LO cross sections for squark-pair production in the infinite
gluino mass limit $\mg \to \infty$ \cite{KL-82}. For finite values
$\mg\;\simlt\; 1\,\mbox{TeV}$, $t$-channel gluino exchange
significantly contributes to the partonic squark cross section: taking
$m_{\tilde g} = m_{\tilde q} = 200$~GeV, ${\tilde u}/{\tilde d}$ pair
production at the Tevatron is enhanced by almost an order of magnitude
as compared to scalar leptoquark pair production.  However, the squark
solutions of the \mbox{HERA} excess events discussed in the literature
belong to the second and third generation \cite{DRR-97,DKM-97}.  The
corresponding pair production cross sections at the Tevatron are
virtually independent of the gluino mass since the $t$-channel gluino
exchange contributions are convoluted with the strongly suppressed
heavy $c,b,t$ parton distributions in the proton.  The ${\tilde c},
{\tilde b}, {\tilde t}$ cross sections are thus completely dominated
by light $u,d$-quark fusion and coincide numerically with those for
scalar leptoquarks, for any value of the gluino mass.

\section{QCD corrections to scalar leptoquark pair production\label{sec-nlo}}
% ------------------------------------------------------------------------- %
Given the potentially large vector leptoquark cross section of
${\cal{O}}(1-10\,\mbox{pb})$ and the absence of a leptoquark signal at
the Tevatron, vector leptoquarks are at best considered as only a
marginal consistent explanation of the HERA data. Moreover, the excess
over the Standard Model expectation at HERA is prominent at large
values of the DIS variable $y$, while vector leptoquark production
would lead to a $y$ distribution $\propto (1-y)^2$.  The most powerful
competitor in this scenario is thus pair production of scalar
leptoquarks. The leading order prediction based on the partonic cross
sections (\ref{eq-ssb}) exhibits a steep and monotonic dependence on
spurious parameters, {\em i.e.}\ the renormalization and factorization
scales: Changing the scales from $\mu=2\ms$ to $\mu=\ms/2$, the LO
cross section varies by $\sim$~100\%.  A refinement of the theoretical
analysis by inclusion of higher-order QCD corrections is thus
mandatory to extract reliable mass limits from the Tevatron
data.\footnote{Soft gluon corrections to the production of leptoquark
  pairs have been discussed in \cite{MM-90} [based, though, on
  erroneous Born calculations].}

\begin{figure}[ht]
\begin{center}
\epsfig{file=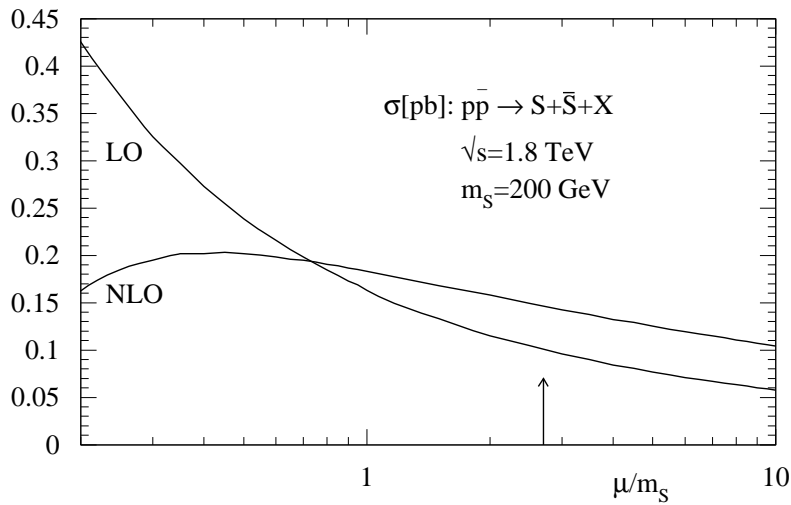,width=10.5cm}
\end{center}
\vspace*{-7.5mm}
\caption[dummy]{\it Renormalization/factorization
  scale dependence of the cross section $\sigma(p+\bar{p}\to
  S+\Slqb+X)$ at the Tevatron energy $\sqrt{s} = 1.8$~TeV.  The arrow
  indicates the average invariant energy $\langle\hat{s}\rangle^{1/2}$
  in the hard subprocess.}
\label{scale-dependence}
\end{figure}

The complete calculation of the next-to-leading QCD corrections has
been performed recently in \cite{KPSZ-97}.  The scale dependence of
the theoretical prediction is reduced significantly when higher order
QCD corrections are included. This is demonstrated in
Figure~\ref{scale-dependence} where I compare the
renormalization/factorization scale dependence of the total cross
section at leading and next-to-leading order. For a consistent
comparison of the LO and NLO results, all quantities [{\it i.e.}\ 
$\alps(\mu^2)$, the parton densities, and the partonic cross sections]
have been calculated in leading and next-to-leading order,
respectively. The NLO cross section runs through a broad maximum near
$\mu\sim \ms/2$, which supports the stable behavior in~$\mu$.

%%%%%%%%%%%%%%%%%%%%%%%%%%%%%%%%%%%%%%%%%%%%%%%%%%%%%%%%%%%%%%%%%%%%%
%%%%%%%%%%%%%%%%%%%%%%%%%%%%%%%%%%%%%%%%%%%%%%%%%%%%%%%%%%%%%%%%%%%%%%

\begin{figure}[ht]
\begin{center}
\epsfig{file=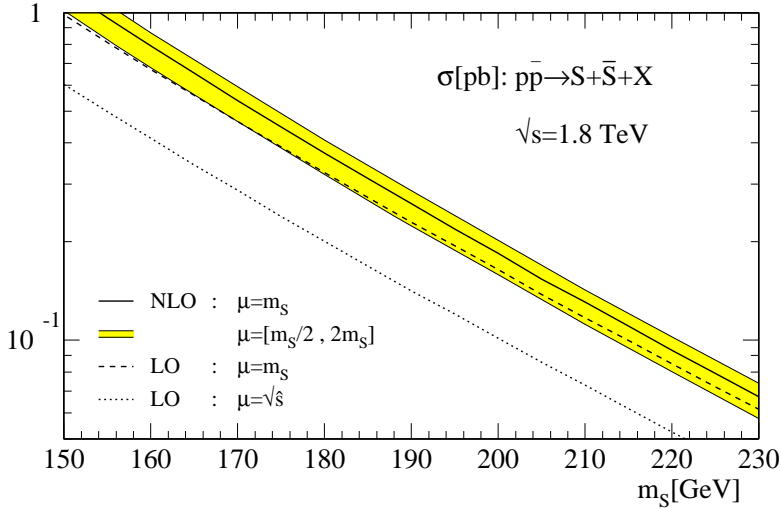,width=10.5cm}
\end{center}
\vspace*{-5mm}
\caption[dummy]{\it The
  cross section for the production of scalar leptoquark pairs,
  $p+\bar{p}\to S +\Slqb+X$, at the Tevatron energy $\sqrt{s} =
  1.8$~TeV as a function of the leptoquark mass $\ms$.  The NLO result
  is compared with LO calculations.  The variation of the NLO cross
  section with the value of the renormalization/factorization scale is
  indicated by the shaded band. The CTEQ4 parton densities have been
  adopted.}
\label{cross-section}
\vspace*{-5mm}
\end{figure}

The QCD radiative corrections enhance the cross section for the
production of scalar leptoquarks above the central value $\mu \sim
\ms$.  If the LO cross section is calculated at large scales $\mu \sim
\sqrt{\hat{s}}$, the enhancement in NLO is as large as $\sim 70\%$,
nearly independent of the leptoquark mass, see
Figure~\ref{cross-section}.  The convergence of the perturbative
approach should however be judged by examining a properly defined
$K$-factor, $K = \sigma_{\mbox{\scriptsize NLO}} /
\sigma_{\mbox{\scriptsize LO}}$, with all quantities in the numerator
and denominator calculated consistently in NLO and LO, and evaluated
at the central scale $\mu = \ms$.\footnote{ It is not legitimate to
  use $\mu = \sqrt{\hat{s}}$ beyond LO since this choice of scale
  results in an error of order $\alps$, no matter how accurately the
  hard-scattering cross section is calculated \cite{CSS-89}.}  In the
interesting mass range between $150 \leq \ms \leq 250$~GeV, these
$K$-factors vary only between 1.20 and 1.08 \cite{KPSZ-97}. They are
small enough to ensure a reliable perturbative expansion.  Since the
cross section for $\mlq\;\simgt\; 150$~GeV is built up mainly by the
quark--antiquark channels, thus based on well-measured parton
densities, the variation between different parton parametrizations is
less than 5\%.

%%%%%%%%%%%%%%%%%%%%%%%%%%%%%%%%%%%%%%%%%%%%%%%%%%%%%%%%%%%%%%%%%%%%%
%%%%%%%%%%%%%%%%%%%%%%%%%%%%%%%%%%%%%%%%%%%%%%%%%%%%%%%%%%%%%%%%%%%%%  

\section{Experimental searches and mass bounds\label{sec-bounds}}
\begin{figure}[t]
\begin{center}
\epsfig{file=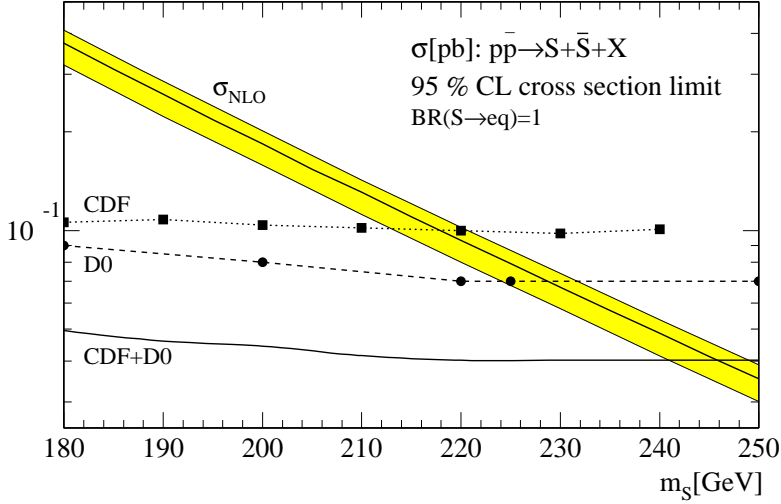,width=10.5cm}
\end{center}
\vspace*{-5mm}
\caption[dummy]{\it The 95\% confidence level upper cross section 
  limits on scalar leptoquark pair production ($\beta=1$)
  \cite{D0-97,CDF-97} compared to the NLO theoretical prediction (see
  Figure~\ref{cross-section}) as a function of the leptoquark mass
  $\ms$. Also shown is an estimate for a combined \mbox{D0} and
  \mbox{CDF} exclusion limit.}
\label{mass-bounds}
\vspace*{-5mm}
\end{figure}

Leptoquark searches have been performed by the \mbox{D0} and
\mbox{CDF} collaborations at the Fermilab Tevatron.  No leptoquark
signal has been found in the full Run~I data sample, resulting in
upper cross section limits for scalar leptoquark pair production.
Figure~\ref{mass-bounds} shows the most recent \mbox{D0} \cite{D0-97}
and \mbox{CDF} \cite{CDF-97} 95\% confidence level limits on the
production cross section times $\beta^2$, where $\beta$ is the
branching fraction of the leptoquark to a charged first generation
lepton plus quark. Comparing the experimental limits with the NLO
theoretical cross section prediction one obtains a lower bound on the
leptoquark mass as a function of the branching fraction $\beta$.
Assuming $\beta=1$ one finds:
\begin{equation}\label{mass-limits1}
\ms \ge \begin{cases} \; 210~\mbox{GeV}\quad \text{[CDF]}\\
                      \; 225~\mbox{GeV}\quad \text{[D0]}
        \end{cases}
\quad \text{for} \quad \beta = 1 \quad (95\%\;\text{C.L.})
\end{equation}
Not taking into account (the small) correlated uncertainties, one can
derive an estimate for a combined D0 and CDF exclusion
limit\footnote{To extend the CDF results beyond 240~GeV, I assume the
  experimental cross section limit to be mass independent in that
  region.}, resulting in
\begin{equation}\label{mass-limits2}
\ms\;\simgt\; 240~\mbox{GeV}\quad \text{[CDF+D0]}
\quad \text{for} \quad \beta = 1 \quad (95\%\;\text{C.L.})
\end{equation}

%%%%%%%%%%%%%%%%%%%%%%%%%%%%%%%%%%%%%%%%%%%%%%%%%%%%%%%%%%%%%%%%%%%%%
%%%%%%%%%%%%%%%%%%%%%%%%%%%%%%%%%%%%%%%%%%%%%%%%%%%%%%%%%%%%%%%%%%%%%  

The limit (\ref{mass-limits2}) can be translated
into an upper limit on the branching fraction $\beta$ of a scalar
leptoquark with mass $\ms = 200$~GeV:
\begin{equation}\label{beta-limits1}
\beta \;\simlt\; 0.5 \quad \text{[CDF+D0]}
\quad \text{for} \quad \ms = 200~\text{GeV}
\end{equation}

The corresponding limits on the masses and branching fractions of
{\em first generation} squarks in supersymmetric theories with $R$-parity
breaking are in general significantly stronger, depending in detail on
the value of the gluino mass (see Section~\ref{sec-tev}). Taking $\mg
= m_{\tilde u/\tilde d} = 200$~GeV for illustration, one finds
\begin{equation}\label{beta-limits2}
\beta \;\simlt\; 0.2 \quad \text{[CDF+D0]}
\quad \text{for} \quad m_{\tilde u/\tilde d} = 200~\text{GeV}
\end{equation}
However, the squark solutions of the \mbox{HERA} excess events
discussed in the literature belong to the second and third generation
\cite{DRR-97,DKM-97}.  As discussed in Section~\ref{sec-tev}, the pair
production cross sections and mass bounds for ${\tilde c}, {\tilde
  b}$, and ${\tilde t}$ squarks are virtually independent of the
gluino mass and coincide numerically with those for scalar
leptoquarks.

It is worth pointing out that the limits on the branching fraction
$\beta$ (\ref{beta-limits1},\ref{beta-limits2}) are conservative in
the sense that no assumptions have been made about the nature of
possible additional decay modes. For leptoquarks/squarks decaying into
electron-neutrino plus quark, additional experimental information
\cite{D0-97} can be used to further strengthen the exclusion bounds.

\section{Conclusions\label{sec-con}}
The leptoquark searches at the Fermilab Tevatron lead to a lower mass
bound of about
\begin{equation}
\ms\;\simgt\; 240~\mbox{GeV}\quad \text{[CDF+D0]}
\quad \text{for} \quad \beta = 1 \quad (95\%\;\text{C.L.})
\end{equation}
for a leptoquark-type scalar particle decaying solely to a first
generation charged lepton plus quark. These results exclude the
interpretation of the excess events found at \mbox{HERA} as being due
to the production of a leptoquark state with chiral and
family-diagonal couplings to fermions.\footnote{Note however that the
  limits are not completely model-independent -- they have been
  derived assuming the particle content of the Standard Model plus one
  additional scalar leptoquark--type state.} The Tevatron bounds are
weakened by suppressing the branching fraction $\beta$ into $(eq)$
final states.  Branching fractions $\beta < 1$ are expected for
squarks in supersymmetric theories with $R$-parity breaking.  The
limits on the masses and branching fractions of {\em first generation}
squarks are in general significantly stronger than those for scalar
leptoquarks, depending on the value of the gluino mass, as discussed
in Section~\ref{sec-bounds}.  The cross sections and mass bounds for
the {\em second} and {\em third generation} squarks ${\tilde c},
{\tilde b}, {\tilde t}$ are however not sensitive to gluino exchange
contributions and identical to those for scalar leptoquarks.

\small
\vspace*{-5mm}
\section*{Acknowledgments}
\vspace*{-3mm}
It is a pleasure to thank Herbi~Dreiner, Tilman~Plehn, Michael~Spira
and Peter~Zerwas for their collaboration and comments on the
manuscript. I have benefitted from discussions and communications with
Carla~Grosso-Pilcher, John~Hobbs and Greg~Landsberg.
\vspace*{-5mm}
%\normalsize

%%%%%%%%%%%%%%%%%%%%%%%%%%%%%%%%%%%%%%%%%%%%%%%%%%%%%%%%%%%%%%%%%%%%%%%%

%%%%%%%%%%%%%%%%%%%%%%%%%%%%%%%%%%%%%%%%%%%%%%%%%%%%%%%%%%%%%%%%%%%%%%%%

\end{document}